\documentclass[3p]{elsarticle}

\usepackage{amsmath}
\usepackage[frame]{xypic}

\usepackage{listings}
\usepackage{acronym}
\acrodef{OOL}{object-oriented language}
\acrodef{OOP}{object-oriented programming}
\acrodef{STL}{standard template library}
\acrodef{MD}{molecular dynamics}

\title{\glsim: a general library for numerical simulation}

\author{Tom\'as S.\ Grigera\corref{inifta}}

\address{Instituto de Investigaciones Fisicoqu\'imicas Te\'oricas y
  Aplicadas (INIFTA) and Departamento de F\'\i{}sica, Facultad de
  Ciencias Exactas, Universidad Nacional de La Plata and CONICET La Plata,
  Consejo Nacional de Investigaciones Cient\'\i{}ficas y T\'ecnicas
  (Argentina)}

\cortext[inifta]{Mailing address: INIFTA, c.c.\ 16, suc.\ 4, B1904DPI
  La Plata, Argentina}

\def\E{\mathbf{E}}
\def\e{\mathbf{e}}
\def\x{\mathbf{x}}
\def\X{\mathbf{X}}
\def\cp{\mathbf{\gamma}}
\def\O{\mathbf{O}}

\def\glsim{\texttt{glsim}}

\def\pseudo#1{{\emph{#1};}}

\begin{document}

\begin{abstract}
  We describe \glsim, a C++ library designed to provide routines to
  perform basic housekeeping tasks common to a very wide range of
  simulation programs, such as reading simulation parameters or
  reading and writing self-describing binary files with simulation
  data. The design also provides a framework to add features to the
  library while preserving its structure and interfaces.
\end{abstract}

\maketitle

\section{Introduction}

Numerical computation or numerical processing of data with digital
computers is an ever-increasing part of day-to-day scientific work,
from statistical analysis of experimental data to numerical solution
of systems of nonlinear integral or differential equations, to large
scale simulations of many-particle systems (from electron gases to
galaxies to social systems \cite{philo:hartmann96}). Many
packages and libraries, both commercial and open-source, exist that
can perform, or aid in performing, a wide range of more or less
specialized tasks. However, due to the very nature of scientific
endeavor, existing software is not always useful for the task at
hand. New problems are studied which require variations or
combinations of known techniques, or new techniques or solutions are
developed for old problems. Thus the practising scientist often finds
him or herself writing computer code.

One of the difficulties faced is that a useful working program
requires much more code than the lines needed for the main algorithm,
due to the need for user interaction, data input/output, and possibly
data preprocessing or format conversion. For example, the code for the
LBFGS algorithm for minimization of a function of many variables
\cite{algorithm:liu89} amounts to about 500 FORTRAN lines, excluding
comments\footnote{As implemented by J.~Nocedal, available from Netlib
  \cite{netlib}.}. A fully working program to find potential energy
minima of a particle system based on LBFGS put together by the author
included an additional 300 lines for user interaction and interfacing
between the energy routines and LBFGS plus 300 for input/output of
simulation trajectory files, plus the routines for evaluation of the
energy. This additional code is ``clerical'': it expresses relatively
simple tasks and straightforward algorithms, and generally takes a
small fraction of execution time. But it must be written, debugged and
maintained alognside the core of the program. When flexibility is
added to the program, the clerical code typically grows
quickly. Maintenance and debugging effort grows quickly with size
unless the code is well structured \cite{software-devel:banker93}, but
well structured code requires thought and planning. Either way,
clerical code ends up requiring a fair amount of attention.

This issue can be more or less sidestepped by writing very rigid,
user-unfriendly software (e.g.\ Monte Carlo programs needing
recompilation to change the temperature), which are highly unlikely to
be useful to anyone but the original author in the original
situation. Though such disposable software may sometimes make sense,
most of the time a little more foresight is desirable. Especially
because if the program is disposable, so tend to be the data it
produces: ad-hoc file formats difficult or impossible to read without
access to the source code that created them.

Ideally, the scientist needing to program some new algorithm should be
able concentrate in writing and debugging the code for the main
algorithm (on which he/she is an expert), while resorting to some sort
of library for the clerical (but indispensable) tasks which are not
part of the main algorithm, and likely not within the scientist's main
expertise.

But can such a general library be developed, or even defined? Not if
the ``algorithm'' and the ``clerical'' tasks remain so vaguely
described.  But we show below that one can define a simulation program
very generally yet precisely, in a way that allows to identify the
basic administrative tasks such a program will need, and build a library to
perform such tasks. We also describe a particular implementation of
such library, called \glsim, which is available for download under an
open-source license (see sec.~\ref{sec:conc}).

It turns out, not surprisingly, that algorithms can be designed
more generally the more abstractly the problem to be solved is
defined.  We thus begin (sec.~\ref{sec:sim}) giving an abstract
definition of a simulation, and listing a series of features that
should be included in a good simulation program. This allows us to
write an outline simulation algorithm (sec.~\ref{sec:algo}). After
commenting briefly on the programming techniques most useful for
designing a library of the kind we are after
(sec.~\ref{sec:techniques}), we describe the \glsim\ library in
sec.~\ref{sec:glsim}. We conclude in sec.~\ref{sec:conc}.

\section{Definition of a simulation}

\label{sec:sim}
\subsection{A simulation: an abstract view}

Take a molecular dynamics or Monte Carlo simulation, or an
optimization technique (conjugate gradient minimization, annealing,
genetic algorithm), or an iterative solution of a system of
differential equations. All of these have in common that they start
from a set of numbers and ``evolve'' this set according to some rules.
The full history of the evolution (``trajectory'') may or not be
interesting in itself, but this is not relevant. The point is that
although very different in aims, these (and other) simulations can be
described under a common scheme.

We shall define \emph{simulation} quite generally as the repeated
application of a transformation to a set of numbers. Let's define two
spaces $\cal{X}$ and $\cal{E}$, which we can assume to be subsets of
$\Re^n$. $\cal{X}$ is the \emph{configuration space}, and a vector $\x
\in {\cal X}$ is a \emph{configuration}. ${\cal E}$ is the
\emph{environment space} and $\e \in {\cal E}$ is an
\emph{environment.}

To perform a \emph{simulation step} means to apply the transformations
\begin{align}
  \e_{n+1} &= \E(\e_n), \label{eq:env-upd} \\
  \x_{n+1} &= \X(\x_n,\e_{n+1}). \label{eq:conf-upd}
\end{align}
The configurations and environments thus form an ordered sequence. We
can define a \emph{simulation time} $t(n)$ through any monotonically
increasing function of the number of steps $n$. The separation into
configuration and environment is somewhat arbitrary, but note that
while $\x_{n+1}$ can depend on $\e_{n+1}$, $\e_{n+1}$ is always obtained
independently of $\x_n$.

The ordered pair $(\x_n,\e_n)$ is the \emph{state} of the simulation
at step $n$ (or time $t(n)$). To start the simulation, we must specify
the initial state $(\x_0,\e_0)$. This state can be constructed from
another real vector $\cp$ through
\begin{eqnarray}
  \e_0&=&\E_0(\cp),\\
  \x_0&=&\X_0(\cp).
\end{eqnarray}
The components of $\cp$ are called \emph{control parameters.}

As the simulation progresses, it may be useful or convenient to
compute subsidiary quantities, called \emph{observables.} along the
simulation. These quantities depend only on the configuration, and
their value is not used at all in computing the successive environment
or configuration, so that their computation can be omitted without
changing the final state. We note them $\O_i(\x_n)$. To define the
observable, a number of parameters will in general be needed, and
these could in principle also evolve, so there will be an environment
associated with each observable. In practice it is often convenient to
merge these environments with the main simulation environment, and we
do so below; the important point is that the environment variables
associated with the observables do not interact with the rest of the
environment.

\subsection{A good simulation program}

A computer program that can iteratively apply the transformations
$\E(\e)$ and $\X(\x)$ is a simulation program. To be deemed a good
simulation program, it should fulfill a number of requirements
(Fig.~\ref{fig:req}).

\begin{figure}[h!]
  \centering
\frame{\parbox{0.8\columnwidth}{
\begin{enumerate}
\item Algorithms of the highest quality
\item Bit-level run reproducibility
\item Invisible run splitting or joining
\item Full human-readable record of simulation conditions
\item Easy user control over simulation parameters
\item Safe early interruption before programmed number of steps
\item Easy continuation after early interruption
\item Minimization of losses due to hardware failure (checkpointing)
\item Ability to read files from earlier versions of the program
\item Easy code maintenance
\end{enumerate}
}}
\caption{Requirements for a good simulation program.}
\label{fig:req}
\end{figure}

Some comments on these requirements.

\begin{enumerate}
\item This is an obvious requirement, the fullfilment of which of
  course depends on the particular technique being coded. However, a
  general advise is to write oneself the algorithms on which one is an
  expert (or close to it), and borrow the rest. This means using good
  libraries. Many are available freely over the internet (e.g.\ Boost
  \cite{boost}, the GNU Scientific Library\cite{gsl,gsl:galassi09}, the Netlib
  collection \cite{netlib}).

\item Computers are good at doing exactly the same things when given
  the same data, so this is not very difficult to achieve. The point
  is to stress that the user must have a way to completley specify
  \emph{all} initial conditions, some of which may not be apparent at
  first sight (like the internal state of the random number generator,
  for instance), so that a given final state can be reproduced
  bit-to-bit. This kind of reproducibility is on the other hand very
  difficult to achieve across architectures, but this is not very
  often a crucial need.

\item This (together with the previous requirement) is most useful in
  debugging or tracking the origin of anomalous behaviour that might
  manifest itself under particular circumstances. This can be achieved
  by saving the final state as binary data using the machine's
  internal representation, avoiding conversions e.g.\ to ASCII decimal
  numbers. 

\item The program must produce a human-readable log file with all
  relevant information to allow the reproduction of the run. This is
  easy but time-consuming to program. A way to automate the production
  of the log is desirable.

\item Programs that require recompilation to change control parameters
  are all too common. This is unacceptable because it means that part
  of the information required to reproduce a run is buried in the
  executable. Of course there is some common-sense imposed limit on
  what should be parametrizable in an algorithm, but quantities that
  are expected to be changed (for tuning or to cover a relevant domain
  of the parameter space) should be stored in a file. For easy user
  control, a text file with an intuitive syntax (\texttt{.ini}-like)
  is preferable. Terminal input is generally not a good idea, as
  programs with a long runtime are likely to be scheduled for remote
  or background execution.

\item If for some reason (like an unexpected need to shutdown the
  machine) the simulation must be interrupted before completion, it
  should be possible to tell the program to save its internal state and
  terminate.  This requires some means of communicating with a
  process which might not be associated with a terminal, such as Unix
  signals.

\item On launching the program again after an early interruption, it
  should automatically recognize a partially completed run and pick up
  from where it left.

\item The above saving of the internal state could be performed
  automatically every few hours, so that the simulation can resume
  with minimal loss of CPU time after hardware failure or power
  outage.

\item When the algorithm is improved or new features are added, it
  often becomes necessary to incorporate new data into the state
  files. These data will not be available when starting a simulation
  from a state written by an earlier version of the program, but the
  new version should be able to read the older files and supply
  appropriate default values for the missing data. This needs the use
  of files with self-describing structure.

\item The code should be organized in a way that it is easy to
  understand, debug and extend, using a modular design. Good
  documentation of source code is essential.

\end{enumerate}

Clearly a good simulation program implementing the above features will
require many lines of code apart from those implementing the specific
algorithm of interest. Desirable though these features are for a
program that will typically run for many hours, implementing them all
is probably out of the question for a small (often one-man)
team. Ideally, one would want a library that allows implementation of
all these requirements automatically or with minimal effort, leaving
the scientist-programmer to concentrate on point 1, specific to
his/her problem.

\subsection{A basic simulation algorithm}

\label{sec:algo}
Taking into account our definitions and requirements, we are in
position to write an outline simulation algorithm
(listing~\ref{lst:algo}).

\begin{lstlisting}[basicstyle=\small,language=Pascal,columns=fullflexible,
emph={endif},emphstyle=\textbf,caption={General simulation algorithm},
label=lst:algo]
read /*@$\cp$@*/
if partially completed run is found on disk then
  read /*@$n, \e_n, \x_n$@*/
else
  create /*@$\x_0, \e_0$@*/ (perhaps from a saved state)
  n=0
endif

repeat
   n=n+1
   compute /*@$\e_n=\E(\e_{n-1})$@*/
   compute /*@$\x_n=\X(\x_{n-1},\e_n)$@*/
   for all observables compute /*@$\O_i(\x_n,\e_n)$@*/
   write /*@$\O_i$@*/
   write log
until /*@$n=$@*/ requested steps or early termination requested

write /*@$\e_n$@*/ and /*@$\x_n$@*/
log termination
end
\end{lstlisting}

The fact that we have been able to say so much about a simulation
without giving any details of the functions $\X(\x)$ and $\E(\e)$
might make us wonder how much of what we have said can be conveyed to
a computer at this level of abstraction. So-called \acp{OOL}
normally offer a set of features that allow us
to code a simulation as so far defined in an actual programming
language, compile the code and build it into a library. Obviously we
will not have functional executable until $\X(\x)$ and $\E(\e)$ are
completely specified and coded, but we shall be close to a ``fill-in
the blanks'' situation where $\X(\x)$ and $\E(\e)$ can be just plugged
in and a full-featured simulation program will result.

\section{Design principles and programming techniques}

\label{sec:techniques}
A more or less obvious requirement for a useful library of the kind we
are after is that it be built in modules with well-defined interfaces,
and with the minimum possilbe interaction among them, so that they can
be plugged in as necessary and combined in different ways. Note
however that a powerful modularization is generally not one based on
processes, or tasks to be performed on some data, but rather one that
represents the division of the problem in abstract parts
\cite{software-devel:parnas72}. These abstract parts will necessarily
embody procedures \emph{and} data: for example, in \glsim\ there is a
module for the concept of environment, which holds data associated
with the environment as well as the procedures to read and write that
data to a file, among others. The interface of a module (how it is
seen from the outside) should reflect the abstraction, while the
implementation details (i.e.\ the design decisions) are kept within
the module. Modules so built are likely to be useful in a wider range
of situations, and keeping design decisions local allows
implementation improvements to be easily integrated into existing
code. Thus modules can be characterized as keeping to themselves one
or more design decisions that they hide from the others. This is known
as \emph{information hiding} \cite{software-devel:parnas72}.

The bundling together of data of different types and procedures
operating on these data is called \emph{encapsulation}
\cite{essay:berard92a}. If several
instances of the data so aggregated can be (easily) created by the
user, these instances are usually called \emph{objects} (sometimes
described as ``intelligent data''). In \acp{OOL}, objects are
variables of a user-defined type, thus the procedure of encapsulation in an
\ac{OOL} amounts to the creation of a new datatype. Languages that
include syntactical support for encapsulation typically offer
facilities to enforce information hiding to some degree, by allowing
to make some data and procedures inaccessible from outside the
module where they are defined. Information hiding
requires encapsulation at the logical design level (even if not
supported syntactically by the language), and perhaps for this reason
the two terms are often used interchangeably \cite{essay:berard92a}.

Another requirement is that it be easily possible to refine, or
\emph{specialize} the modules provided by the library, adding to them
new capabilities without breaking the interface.  Adding data and
procedures to an existing module or object ``specializes'' the object
in the sense that to add capabilities one must typically make more
assuptions on how the object will be used and/or impose restrictions
on the operations allowed: the specialized object has additional, more
specific properties \cite{review:taivalsaari96} and thus represents a
less general concept. In \acp{OOL}, it is generally convenient to
implement specialization through the mecanism of \emph{inheritance,}
which is the possibility to define a datatype based on
an existing datatype, only defining explicitly the properties desired
for the former that differ from those of the latter
\cite{review:taivalsaari96}. Note however that inheritance and
specialization are not isomorphic concepts, and that there are uses of
inheritance other than specialization \cite{review:taivalsaari96}.

A bit more subtly, our insistence on building the library as far as
possible around abstract concepts requires in turn to be able to write
algorithms abstractly, or generically, in the sense that it must be
possible to include in the library modules using undefined procedures
and/or operating on unspecified data types or data types not
completely defined.  For this \glsim\ relies heavily on
\emph{polymorphism.}

Polymorphism \cite{cs:strachey00} refers to the ability of handling
different data types with a uniform interface, which can also be
described as using a single name to call different functions, based on
the types of the data to be passed to that function. This includes a
wide variety of situations.  \emph{Ad-hoc} polymorphism
\cite{review:cardelli85} refers to the case where different data types
are processed by calling different functions, given explicitly for
each combination of the allowed types. This includes overloading
(writing many functions of the same name but different argument types)
and coercion (automatic conversion of some types to other
types). Ad-hoc polymorphism
is found to some degree in all common programming languages (for
instance, one uses the symbol $+$ for addition of two integers or two
floating-point variables). For our goal of writing abstract
algorithms, we need a language supporting \emph{universal}
polymorphism \cite{review:cardelli85}, which means that the same code,
or code generated using the same rule, is used for all admisible
types. One way to achieve this is writing functions where types are
not specified but left as a parameter. This is called \emph{parametric
  polymorphism} \cite{review:cardelli85}, or generic programming in
\ac{OOL} jargon. Another, perhaps more powerful, possibility is
\emph{inclusion} (or \emph{subtype}) polymorphism
\cite{review:cardelli85}, afforded by the concept of inheritance:
objects of datatype B derived (i.e.\ defined by inheritance) from
datatype A can be thought as being of type B or type A. Thus code
written to operate on objects of type A can also operate on objects of
type B.

Polymorphism is what allows us to write and compile our algorihtm:
clearly $\X(\x)$ is a name that refers to different (unknown at
compile time) functions, which we can distinguish by looking at the
type of the argument. We shall be using mostly inclusion polymorphism
because, at least in the language of our implementation (C++), it
allows to explicitly express the assumptions one is making about the
type that the function will be handling. Put in another way,
inheritance guarantees a minimum set of operations that can be
performed on the object to be processed.

Substituting different functions for one name is something that has
been possible even in very old languages, by combining different
modules at the linking stage. This can perhaps be thought of as a
rather primitive substitute for polymorphism, as one could compile a
module coding the basic algorithm and substitute the appropriate
$\X(\x)$ at link time. Indeed, the design of \glsim\ profits from
experience gained in developing a modular system written in C using
such a scheme. However, this is not polymorphism, since the function
is not selected based on argument type, but arbitrarily, outside the
language itself. It thus cannot benefit from language features such as
type checking. Also, static binding (i.e.\ mapping names to functions
at compile or link time) has limitations, since it is not always
possible to know at compile- or link time which function it is desired
to call (which can be alternatively expressed by stating that the type
of the argument is not always known at compile time). For instance,
when linking a simulation program one typically wants to include only
one of the possible $\X(\x)$ functions. However, the situation is
different with observables. Similarly to the simulation algorithm, it
is best to write some generic code for the observation only once (see
sec.~\ref{sec:observables}) and leave to the user just the task of
writing code for the specific quantity required, by supplying some
function $\O(...)$. Since one may want more than one observable
computed during the same simulation, the generic code will need to
call different $\O(..)$ functions at different times in the same point
of the program. This requires \emph{dynamic binding,} or the ability
to select the appropriate function at run time. Dynamic binding is
necessary to exploit the full power of inclusion polymorphism.

\subsection{Object-oriented programming and C++}

So-called \acfp{OOL} provide syntax constructs that allow to easily
express the techniques mentioned above. ``Easily'' means that concepts
such as inheritance (which could conceivably be used in a program
written in, say, C) can be expressed in the syntax of the language and
thus more conveniently, with less work on the part of the programmer,
and in a way such that the compiler ``understands'' what the
programmer is trying to do and can help with compile time checks and
diagnostics (e.g.\ type checking).

\glsim\ is written in C++ \cite{book:stroustrup97}, which is a
standarized language with static typing that supports encapsulation,
inheritance and polymorphism, and for which compilers are available in
a wide variety of plattforms. In C++ encapsulation is achieved by
defining \emph{classes,} which are user-defined datatypes. The
procedures bundled with the data within a class are called
\emph{methods} or \emph{member functions,} and data and procedures can
be classified as public or private, in which case are inaccessible
from functions defined outside the scope of the class. Instances of a
class are called \emph{objects.}  Polymorphism is supported through
explicit overloading (ad-hoc), \emph{templates} (parametric), and
inheritance (inclusion). Binding is static by default, but
dynamic binding can be requested for specific methods by declaring
them \emph{virtual.}

When a method is virtual, C++ allows the programmer to declare it but
leave it undefined (i.e.\ not coded). These methods are called
\emph{pure virtuals.} Classes with pure virtual methods cannot be
instantiated, because the compiler would not know what to do when
these methods are called. They can only be specialized by defining
classes derived from them (at some point the pure virtuals will be
defined and it will be possible to create instances of those derived
classes). For this reason they are called \emph{abstract base
  classes,} or ABCs. Classes like Simulation and Configuration (see
Fig.~\ref{fig:classes}) are ABCs in \glsim. ABCs are useful for
interface specification.

\glsim\ can be said to use \ac{OOP} to the extent that it uses the
techniques we have described \cite{book:stroustrup97,
  book:berard92}. However, OOP is sometimes described as a way to
match ``real-world'' objects to software entities in a way that allows
more convenient manipulation of them for computing purposes. The
author's experience suggests that this view may be misleading, or too
narrow, as a class hierarchy design tends to be more useful when built
around rather abstract concepts, which may be hard to trace to ``real
world'' objects. Also, the language may force or induce the programmer
to define classes to benefit from features such as encapsulation,
resulting in objects that may not be the most intuitive. For example,
\glsim\ defines a class for the simulation. It is hard to make the
case that the simulation itself is a ``real world'' entity. However
the simulation class turns out to be a good programming solution that
allows to use C++ support for dynamic binding.

In summary, we simply claim that \glsim\ is written making use of
encapsulation and polymorphism as techniques for dealing with the
present problem in an abstract way, and is thus written in a
widely-available language with good support for them.

A final word about language choice. C++ is often critised as being
slow (although the criticism is contested
\cite{book:stroustrup97}). It might thus seem a poor choice for a
simulation package. Without entering the speed discussion, let us
simply point out that the clerical code \glsim\ mainly deals with is
not performance-critical. Most of the CPU time will likely be spent
computing the transformation $\X(\x)$ (think of the force loop of
molecular dynamics, for instance). If the programmer deems C++ too
slow for the core part of the simulation, he is free to chose any
other language; interfacing with \glsim\ will still be possible in
reasonable platforms. While \glsim\ strives of course to be as
efficient and fast as possible, it is clear that the (likely small)
performance penalty introduced will be more than offset by the savings
in expert human time required to produce a good simulation program.

\subsection{Templates vs.\ virtual functions}

As we said, \glsim\ relies heavily on universal polymorphism. In C++
this means using templates (parametric polymorphism) or virtual
functions (subtype polymorphism). The advantages of templates are that
they produce slightly smaller objects (because objects with virtual
functions need to store a table of virtuals, the so-called vtable),
and that they result in faster code, because virtual functions, being
resolved at run-time, are called via an indirection. In particular,
virtual functions cannot be inlined.

Polymorphism through templates should thus be preferred where speed is
critical. If the function implementing the transformation $\X(\x)$
relies on polymorphism, it probably should use templates. \glsim\
however uses virtual functions because when implementing
non-performance-critical tasks polymorphism through virtual functions
has the following advantages over templates:
\begin{enumerate}
\item It allows explicit, compiler-checkable interface
  specification: by writing ABCs, the pure virtuals make
  explicitly obvious what functions the user of the class expect to
  find implemented.
\item It allows for dynamic polymorphism: all objects of a derived
  type can be accessed through pointers to the base type. One can
  thus make containers that hold objects of different type. This
  feature is used in \glsim\ for instance to deal with observables:
  all observables descend from \lstinline!class Observable! and are
  accessed from the simulation class through a list of pointers to the
  base class (sec.~\ref{sec:observables}). This does not work with
  templates because instantiating a template creates a completely
  different type.
\item Virtual functions stay virtual for ever down the class
  hierarchy: one can specialize a class by deriving and defining or
  overriding the virtuals in the base, then further specialize the
  second class by deriving again and overriding only some of the
  virtuals. This is not easily and conveniently done with templates.
\end{enumerate}

\subsection{Literate programming}

We have said that we would like a program as easy as possible to
mantain and debug, and that this requires among other things a good
documentation of the source code. In an attempt to achieve good and
up-to-date source-code documentation, \glsim\ is written using the
\emph{literate programming} style of programming
\cite{book:knuth92}. The idea is to write the program and
documentation simultaneously, shifting the focus from instructing a
computer what to do to explaining to a human being what we want the
computer to do \cite{book:knuth92}. The result should be a sort of
``essay'' that combines code and documentation.

In practice, the programmer writes a file containing documentation
chunks written in a text-formatting language (in our case \LaTeX) and
code chunks written in some programming language (C++ in our case). A
literate-programming tool is needed that on one hand extracts the
source code and prepares a file suitable for the compiler, and on the
other adds the necessary formatting commands to produce a \LaTeX\ 
source file. \glsim\ uses \texttt{noweb,} a freely-available literate
programming tool \cite{software-devel:ramsey94}. After processing with
\LaTeX, the result is a document (Fig.~\ref{fig:literate}) that reads
roughly like the description given below, except of course that all
details and the full source are included.

\begin{figure}
\center
\includegraphics*[width=0.5\columnwidth]{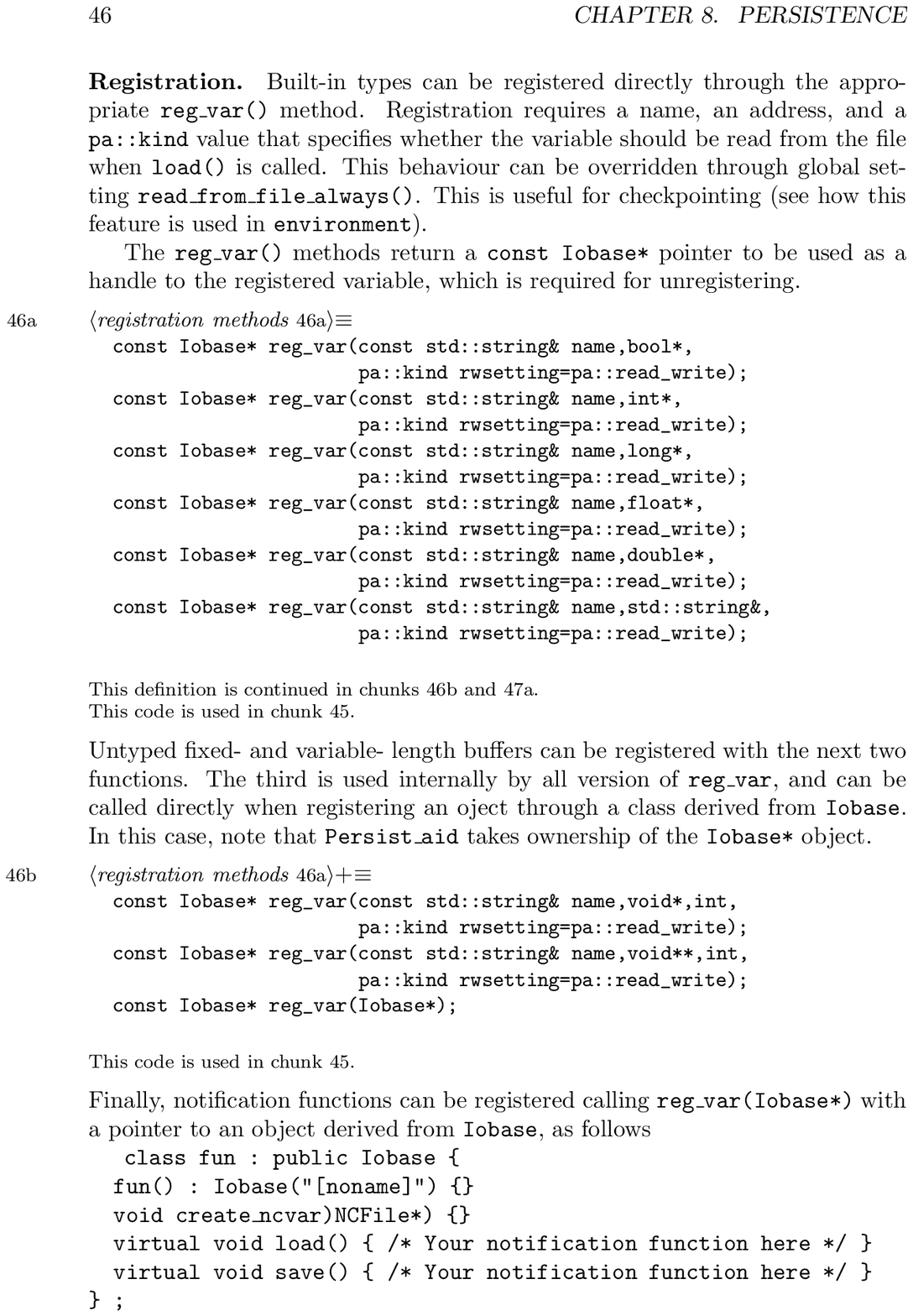}
\caption{The documented source after processing with \texttt{noweb}
  and \LaTeX.}
\label{fig:literate}
\end{figure}

\section{An overview of the \glsim\ library}

\label{sec:glsim}
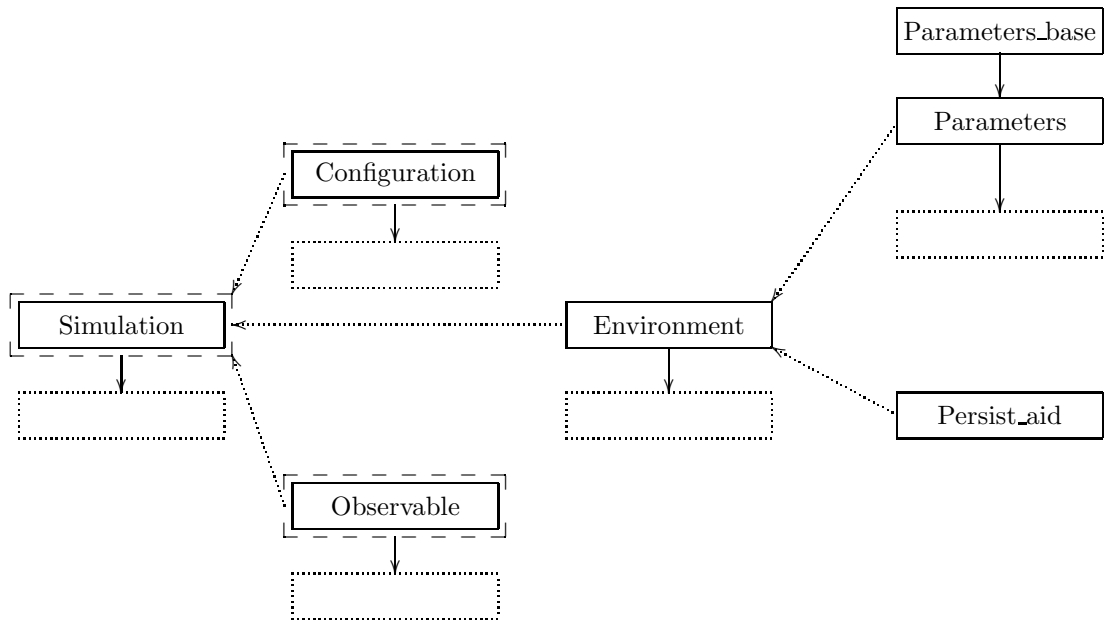
\begin{figure}
\centering
\begin{xy}
\def\aclass#1{\xybox{*=(0.9,0.2)\txt{#1}*\frm{-}*+\frm{--}}}
\def\class#1{\xybox{*=(0.9,0.2)\txt{#1}*\frm{-}}}
\def\emptydbox{\xybox{*=(0.9,0.2)\frm<10pt>{.}}}
\POS
0 ;  <3cm,0cm> :                        
(1,1)*\aclass{Simulation};p="sim"        
    ,p+D;-(0,0.4)*\emptydbox{}+U**\dir{-}*\dir{>}
,"sim"+D+(1.2,0.8)*\aclass{Configuration};p+L="conf"  
    ,p+D;-(0,0.4)*\emptydbox{}+U**\dir{-}*\dir{>}
,"conf";"sim"+RU**\dir{.}*\dir{>}                     
,"sim"+(2.4,0)*\class{Environment};p="env"
    ,p+D;-(0,0.4)*\emptydbox{}+U**\dir{-}*\dir{>}
,"env"+L;"sim"+R**\dir{.}*\dir{>}
,"sim"+(1.2,-0.8)*\aclass{Observable};p+L="obs"
    ,p+D;-(0,0.4)*\emptydbox{}+U**\dir{-}*\dir{>}
,"obs";"sim"+RD**\dir{.}*\dir{>}
,"env"+RU+(1,0.8)*\class{Parameters};p="par"
     ,p+(0,0.4)*\class{Parameters\_base}
     ,+D;+U**\dir{-}*\dir{>}
     ,"par"+D;p-(0,0.4)*\emptydbox{}+U**\dir{-}*\dir{>}
,"par"+L;"env"+RU**\dir{.}*\dir{>}
,"env"+R+(1,-0.4)*\class{Persist\_aid}
+L;"env"+RD**\dir{.}*\dir{>}
\end{xy}
\caption{Class hierarchy diagram. Full arrows indicate inheritance and
  dotted arrows composition (the class pointed by the arrow includes a
  pointer or reference to an instance of the other class).  Dashed
  borders indicate abstract classes; empty dotted boxes represent
  classes the user should define, inheriting from the library as
  shown, in order to produce a functional simulation program.}
\label{fig:classes}
\end{figure}

Let us give an overview of the library from the point of view of the
user. We shall leave out many details and show some parts as
pseudo-code, so that to actually write code using the library it will
be necessary to read the documentation accompanying the
package. However the present description should give the reader an
idea of the internal organization, and of what the user can expect
from \glsim.

\glsim\ uses classes to represent the concepts of configuration,
environment and observable introduced in sec.~\ref{sec:sim}. It also
turns out to be useful to introduce a simulation class implementing
our main algorithm (listing~\ref{lst:algo}) in one of its
methods. This method is virtual so that it can be eventually
overridden. With the exception of \lstinline!class Environment!, these
classes are abstract, because they include at least one pure virtual
function. A few additional classes are defined to read the
configuration file (\lstinline{class Parameters}) and to automate the
production of self-describing files
(\lstinline!class Persist_aid!).
Fig.~\ref{fig:classes} gives an overview of the class
structure.

If the user wants to write, say, a Monte Carlo simulation of the Ising
model, (s)he would inherit from Configuration to declare an
appropriate spin lattice, inherit from Simulation to write the
Metropolis sweep, and from Environment to add the necessary
parameters, such as temperature and coupling, and optionally define
one or more observables inheriting from \lstinline{class Observable.} To
run the simulation, the main program then simply creates the objects,
tells the configuration and environment to initialize themselves from
disk and calls the Simulation run method. On completion of the run,
the configuration and environment save methods are
called. \lstinline{main()} would read something like the following.

\begin{lstlisting}[caption={Simulation main()},label=glsim-main]
extern Parameters *create_parameters();
extern Environment *create_environment(int argc,char *argv[],Parameters *par);
extern Configuration *create_configuration(Environment *env);
extern Simulation *create_simulation(Environment*,Configuration*);

int main(int argc,char *argv[])
{
  Parameters    *par=create_parameters();
  Environment   *env=create_environment(argc,argv,par);
  Configuration *conf=create_configuration(env);
  /*@\pseudo{load configuration}@*/
  Simulation *sim=create_simulation(env,conf);
  /*@\pseudo{load environment}@*/
  steps=sim->run();
  /*@\pseudo{save environment and configuration}@*/
  delete sim;
  delete conf;
  delete env;
  delete par;
}
\end{lstlisting}

This main (with the addition of exception catching and timing
functions) is included in \glsim, since its structure should not need
alteration, and since the objects should be created in the order shown
(in particular, the environment must be loaded after the simulation
has been created, because the environment reads all variables
registered for automatic saving, as discussed below). Different
simulations can simply link this main with the appropriate
\lstinline{create_xxx()} functions.

\subsection{Simulation}

The Simulation class implements our abstract simulation algorithm
(Listing~\ref{lst:algo}). A
Simulation object is created by passing pointers to suitable
Configuration and Environment objects. The required observables are
created (defining objects from a separate hierarchy described below,
sec.~\ref{sec:observables}) and registered by calling
\lstinline{add_obs()}. This guarantees that the simulation is aware of
them, and that the methods to compute the observables will be called
when appropriate. Finally, the public run method is called to run the
simulation as shown above. To produce a working simulation, the user
must inherit from \lstinline{class Simulation} and define the functions
\lstinline{init_sim()}, \lstinline{step()}, and the logging functions.

\begin{lstlisting}[caption={Declaration of class Simulation (excerpt)}]
class Simulation {
public:
  Simulation(Environment &e,Configuration &c);
  virtual const char *name()=0;
  virtual long run();
  void add_obs(observer* o);
  
protected:
  virtual ~Simulation();
  virtual void step()=0;
  /*@\pseudo{logging functions}@*/
  /*@\morecode@*/
  virtual void start_observation();
  void obs();
  /*@\morecode@*/
} ;
\end{lstlisting}

The private part holds references to the environment and configuration
objects (secs.~\ref{sec:configuration} and \ref{sec:environment}) and
a list of pointers to the observables. We have omitted the
declarations and functions to install a signal handler for the Unix
termination signals. The handler simply sets the
\lstinline{termination_requested} flag and returns, so that the
current step is completed. The main simulation loop, below, checks
this flag and stops the simulation if set. In this way the simulation
can be safely interrupted with the \texttt{kill} command, or with
\texttt{ctrl+C} if running interactively.

\begin{lstlisting}[caption={Simulation run method}]
volatile sig_atomic_t Simulation::termination_requested=0;

long Simulation::run()
{
  /*@\pseudo{install signal handler}@*/
  check_for_partial_run(); /* This functions sets rmode */
  init_sim(rmode);
  if (env.obs_step>0) start_observation();
  log_start_run();

  env.completed_run=false;
  long steps_completed=env.requested_steps-env.steps_remaining;
  long actual_steps=0;

  /* Simulation loop; if a signal is received the handler will set termination_requested to 1 */
  while (env.steps_remaining>0 && termination_requested==0) {
    env.step();
    step();
    if (env.total_steps%env.log_step==0) log();
    if (env.obs_step>0 && env.total_steps%env.obs_step==0) obs();
    env.steps_remaining--;
    actual_steps++;
  }

  steps_completed+=actual_steps;
  if (termination_requested>0)
    std::cout << "\nWARNING: Terminating on signal " << signal_received <<
      "\nCompleted " << steps_completed << " steps.\n\n";
  else
    env.completed_run=true;
  log_stop_run();
  return actual_steps;
}
\end{lstlisting}

Observables are handled by keeping a list (structure from the
\ac{STL}) of pointers to them,
\lstinline{std::Observable*}~\lstinline{observables}.
 \lstinline{class Observable} (sec.~\ref{sec:observables}) provides methods for
inizialization and observation (i.e.\ computation and saving of the
desired quantities), so that \lstinline!Simulation::obs()! simply goes
through the list, calling the appropriate method for each observable.

\subsection{Configuration}

\label{sec:configuration}
A configuration is required to have a name, to know how to
load and save itself to disk, and to be able to initialize itself to
some default (say, a random but valid configuration).
\begin{lstlisting}[caption={Declaration of class Configuration}]
class Configuration {
public:
  std::string name;
  Configuration(const std::string& name_) : name(name_) {}
  virtual ~Configuration() {}
  virtual void deflt()=0;
  virtual void load(const char*)=0;
  virtual void save(const char*)=0;
} ;
\end{lstlisting}

A working (i.e.\ instantiable) configuration class must define the
three pure virtuals above plus the appropriate access interface,
through which the \lstinline{Simulation::step()} method will update
it. It is important to keep it light, since the configuration will be
accessed and updated many thousands of times during a run. In many
cases, public data members are probably the best alternative. See
sec.~\ref{sec:example} for an example.

If desired, loading and saving in self-describing files can be done
using \lstinline{class Persist_aid} below. However, configurations must
be saved in files physically distinct from environment files.

\subsection{Environment}

\label{sec:environment}
The environment holds all the data relevant to the simulation
which is not reasonable to include in the configuration, including the
number of steps completed and requested for the run, and filenames to
read and save environment and configuration. An environment object can
be created passing it the names of those files, or alternatively the
\lstinline{argc} and \lstinline{argv} arguments of function \lstinline{main}
plus a reference to an object of \lstinline{class Parameters}
(sec.~\ref{sec:parameters}). In this last way, it will parse the
command line and initialize itself from a control file.

The declaration of \lstinline{class Environment} is shown below (many
variables omitted for brevity). Since the data are typically to be
manipulated from outside the class, public data access has been
preferred over get/set methods.
\begin{lstlisting}[caption={Declaration of class Environment (excerpt)}]
class Environment {
public:
  std::string title;
  int         requested_steps;
  int         log_step;
  std::string configuration_file_ini,configuration_file_fin;
  long        total_steps;
  /*@\morecode@*/

  Environment(int argc,char *argv[],Parameters &param);
  virtual ~Environment() {}
  virtual void step();
  void load();
  void load_all();
  void save();

protected:
  Persist_aid  persist;
  /*@\morecode@*/
} ;
\end{lstlisting}

The environment is updated (i.e.\ the action of the function $\E(\e)$
of Eq.~(\ref{eq:env-upd}) is performed) by calling \lstinline{step()}
(which is done from \lstinline{Simulation::run()}). At this level the
only action required is to increment the number of steps
(\lstinline!total_steps!), but more complicated things can be done by
overriding this virtual.

The i/o methods are \lstinline{load()} and \lstinline{save()}, for normal
reading/writing to the environment file, and \lstinline{load_all()},
which is the load function to be called when continuing a previously
interrupted run (see discussion in sec.~\ref{sec:persistence}).
Variables are read and written through a \lstinline{Persist_aid} object
(sec.~\ref{sec:persistence}). \lstinline{class Persist_aid} does i/o to a
self-describing binary file, so that variables are read by name. In
this way it is possible to read old versions of environment files,
because when variables are missing, a warning is printed and a default
value (typically set from the configuration file) is
used. 

\lstinline{Persist_aid} works through a simple registration mechanism
as illustrated in the constructor below, making it easy to incorporate
to the environment file variables defined by the user.  This is done
inheriting from \lstinline{Environment}. If the new variables are to
be initialized from the control file, a class derived from
\lstinline{class Parameters} (sec.~\ref{sec:parameters}) is first
defined which declares the necessary variables to the parameter file
parser. This object is passed to \lstinline!Environment!'s derived
constructor, which reads the parameter file values calling
\lstinline{Parameters::value()}, and registers the
variables to be saved with the \lstinline{Persist_aid} object.

\begin{lstlisting}[caption={Environment constructor (excerpt)},label=lst:env::env]
Environment::Environment(int argc,char *argv[],Parameters &param) : 
  ignore_partial_run(false),
  completed_run(false),
  obs_step(0),
  total_steps(0),
  par(&param)
  /*@\morecode@*/
{
  // Parse the control file
  // N.B. this parses *all* defined variables; must not be  called again by derived constructors
  par->parse(argc,argv);

  // Read values from control file
  ignore_partial_run=par->value("ignore-partial-run").as<bool>();
  title=par->value("title").as<std::string>();
  requested_steps=par->value("steps").as<int>();
  /*@\morecode@*/
  register_vars();
}
/*@\morecode@*/
void Environment::register_vars()
{ // Tell persist which variables must be saved in the environment file
  // (derived classes should register their own variables)
  persist.reg_var("environment.title",title,pa::write_only);
  persist.reg_var("environment.requested_steps",&requested_steps,
                  pa::write_only);
  persist.reg_var("environment.total_steps",&total_steps);
  /*@($\ldots$)@*/
}
\end{lstlisting}

\subsubsection{Parameters}

\label{sec:parameters}
Simulation parameters are read form a control file with a
straightforward ``.ini'' syntax (\texttt{variable=value}). Parsing is
done with the \texttt{program\_options} library, a part of Boost \cite{boost},
which can also do command-line parsing. \lstinline!class Parameters_base!
provides a simple interface to
\lstinline!Boost::program_options!. Basically
\lstinline!Paramters_base! defines an object,
\lstinline!ctrl_file_options!, through which configuration-file
parameters can be defined as shown below, and a method
\lstinline!value(const std::string &parameter)! which returns the value of the
requested parameter as a
\lstinline!Boost::program_options::variable_value! object (see the
\lstinline!Environment! constructor in listing~\ref{lst:env::env} for
sample usage and the Boost documentation~\cite{boost} for details).

To define parameters, the user inherits from \lstinline!class Parameters!:
\begin{lstlisting}[caption={Class Parameters declaration (excerpt)}]
class Parameters : public Parameters_base {
public:
  Parameters();
protected:
  void parse_command_line(int argc,char *argv[]);
  /*@\morecode@*/
} ;
\end{lstlisting}

The parameters to be read must be defined before the file is
parsed. The parser is called from \lstinline{parse_command_line}, which
in turn is called by the Environment constructor. The most convenient place to
define the variables is in the constructor of the class derived from
parameters, like it is done in \lstinline!class parameters! itself:
\begin{lstlisting}
Parameters::Parameters() :  Parameters_base()
{
  ctrl_file_options.add_options()
    ("title",po::value<std::string>()->default_value("[untitled]"),
     "simulation title")
    ("steps",po::value<int>()->default_value(1),"number of steps to run")
    ("log_step",po::value<int>()->default_value(0),"write to log every ... steps")
  /*@\morecode@*/
  ;
}
\end{lstlisting}
Parameters are defined by giving a name, a type, a default value and a
description, using the syntax of the Boost::program\_options library
(\lstinline!ctrl_file_options! is an object of type
\lstinline!Boost::program_options::options_description!).

Command-line parsing is done from \lstinline!parse_commmand_line!,
which is a protected virtual so that it can be overriden if
needed. The version implemented in Parameters recognizes a
command-line of the form
\begin{quote}
 \texttt{simprog [options] control\_file
  initial\_infix final\_infix},  
\end{quote}
where \lstinline!control_file! is the file with the control parameters
and initial and final infix are used to build the input and output
filenames according to the based on the patterns given in the control
file (in the variables \texttt{env\_file}, \texttt{conf\_file} and
\texttt{obs\_file}). The special initial infix \texttt{+++} is
interpreted as a request to generate a default environment and
configuration. The options \texttt{-c} and \texttt{-e} are also
recognized, which allow to override the infix-generated configuration
and environment files, respectively.

\subsubsection{Persistence}

\label{sec:persistence}
Class Persist\_aid is designed to easily implement our requirements 7,
8 and 9, namely to be able to read old versions of simulation files
and to transparently resume execution after early interruption.  The
object to be placed under \lstinline!Persist_aid! management is
registered by calling \lstinline{reg_var}. The user can then essentially
forget about loading or saving: all registered variables are saved and
loaded through \lstinline{Persist_aid::save()} and
\lstinline{Persist_aid::load(),} which are typically called at the end
or start of the simulation by \lstinline{Environment} load or save
methods. Say the user wants to save the temperature in the environment
file to have it automatically restored on resuming the simulation,
(s)he would simply do
\begin{lstlisting}
  double temperature;
  persist.reg_var("my_environment.temperature",&temperature);
  /*@\emph{or}@*/
  persist.reg_var("my_environment.temperature",&temperature,Persist_aid::write_only);
\end{lstlisting}
(the meaning of the second form is explained below). The
``(scope-or-namespace).(variable-name)'' convention is suggested to
help keep variable names unique. If this is done in the constructor of
a class inherited from Environment, then \lstinline!persist! is the
\lstinline!Persist_aid! object defined in Environment. Its load and
save methods are called when appropriate and no further action is
needed. On the other hand, if it is desired to keep these data in a
file separate from the global environment file, a different
\lstinline!Persist_aid! object needs to be created and its save and
load methods called as needed.

The save method writes all the registered variables to a
self-describing binary file (at this time managed throught the NetCDF
library \cite{netcdf}). The self-describing nature of the file means
that on reading, variables are looked up by name (the name given on
registering), rather than based on their position in the file. Thus if
a new version of the program attempts to read a file produced by an
earlier version which used to register fewer variables, the variables
common to both versions will be retrieved by \lstinline!load()!
without problems. The user controls what happens when attempting to
read a variable missing in the file by calling one of the methods
\lstinline{on_absence_ignore}, \lstinline{on_absence_warn} or
\lstinline{on_absence_throw}: the missing variable is silently
ignored, a warning is printed on standard output (in both cases the
variable is not changed, so that if it was initialized to a reasonable
default the simulation can proceed) or an exception is thrown.

There is an additional subtelty on reading: some variables (for
instance, the number of steps completed so far) must be kept during a
run, but once the run is finished and a new run is requested starting
from the previously achieved state, they must be reinitialized to new
values. In principle, they should not be saved with the rest of the
environment. However, if the run is interrupted early, those values
are needed to correctly resume the simulation when required. For this
reason, a third argument can be given to \lstinline!reg_var!, taking
one of the values \lstinline{read_write} (the default) or
\lstinline{write_only}. In the default reading mode
(\lstinline{read_from_file_ad_hoc}), \lstinline{write_only} variable
are not read. When resuming a run, \lstinline{Simulation} calls
\lstinline!Environment::load_all()!, which temporarily sets the read
mode to \lstinline{read_from_file_always}, ensuring that even
variables flagged as \lstinline{write_only} are loaded.

All simple (built-in) types, plus C- and C++-style strings can be
registered. It is also possible to register save and load functions
requiring a pointer to a C or C++ file stream. In this case, the
provided save function is called, the data is placed in a buffer and
it is written as a single variable. This is mainly intended to allow
for the use of third-party libraries that provide read/write methods
of their own. Finally, notification functions can be registered which
are called on i/o on a variable, so that for instance derived
quantities can be recomputed when a variable is read from disk.
  
\subsection{Observables}

\label{sec:observables}
The final component is \lstinline{class Observable}, intented as a base class for objects represnting observables.
\begin{lstlisting}[caption={Declaration of class Observable (excerpt)}]
class Observable {
public:
  Observable(const std::string& name, Environment &e,Configuration &c,int st);
  virtual ~Observable() {}
  virtual void start(Simulation::run_mode rmode);
  void observe();
  virtual void register_for_persistence(Persist_aid&);
protected:
  virtual void do_observation();
  /*@\morecode@*/
} ;
\end{lstlisting}

To produce a working observable object, the user must write (appart
from constructor and desctructor), the methods \lstinline{start()} and
\lstinline{do_observation()} and optionally
\lstinline{register_for_persistence()}. The first of these is passed a
parameter telling it whether it should initialize for a new or a
continuation run. It is expected that this function will open a file
to record the observations, so this information is
important. Typically, in a normal run the file is opened in overwrite
mode and a header is written, while a continuation run requires
opening in append mode, and the header is omitted. After initializing,
the parent \lstinline{start()} must be
called. \lstinline{do_observation()} must do the actual calculation of
the observable, accessing the relevant data through the references to
the environment and configuration stored in the object.

If required, variables can be registered with the persist object by
overriding \lstinline{register_for_persistence()}, but it must be
remembered to call the corresponding method in the parent class.

\subsection{Checkpointing and disk files}

At present \glsim\ still lacks support for checkpointing. The reason
is that the existence of disk files to which information is added as
the simulation proceeds (those produced by \lstinline{class Observable}'s
descendants) make checkpointing a harder problem than continuation
after interruption with a signal. It is fairly easy, using Unix alarm
signals to make the program save the state (configuration and
environment) periodically (say every two or three hours). However, if
the system fails, the observable files will be out of synchronization,
because the observable is typically written more often than the
configuration (writing the configuration after each step is not
feasible because it is in general too expensive). Thus restarting
after system failure requires a way to restore the observable files to
the state they were in at the time the last configuration and
environment were written. A convenient mechanism to do this is still
missing, so at this point checkpointing with \glsim\ is not easily
achievable. Alternative ways to provide this mechanism are being
considered, and it is expected that checkpointing support will be
added in the near future.

\section{Example use of \glsim}
\label{sec:example}

Let us sketch how a user would proceed to write a working simulation
program based on \glsim. Assume one wants to implement a \ac{MD}
simulation of monoatomic particles.

First we need to decide how the state of the system (here the
mechanical state of particles in 3-$d$ space) will be represented. We
then define a suitalbe class, inheriting from \lstinline!Configuration!:
\begin{lstlisting}
class MD_configuration : public Configuration {
public:
  int       N;
  double    time;
  double    box_length[3];

  short     *id;
  double    (*r)[3];
  double    (*v)[3];
  double    (*a)[3];

  olconfig();
  ~olconfig();

  void load(const char* fname);
  void save(const char *fname);
  /*@\morecode@*/
} ;
\end{lstlisting}

The \ac{MD} algorithm will need additional parameters, such as the
integration time step. These would be added inheriting from
\lstinline!Parameters!:
\begin{lstlisting}
class MD_parameters : public Parameters {
public:
  MD_parameters() : Parameters()
  {
    ctrl_file_options.add_options()
      ("deltat",po::value<double>,"integration time step") ;
  }
} ;
\end{lstlisting}

These would be kept in an appropriate \lstinline!Environmnet!
descendant, together with additional information that makes sense to
store, e.g. the energy:
\begin{lstlisting}
class MD_environment : public Environment {
public:
  Environment(int argc,char *argv[],MD_parameters& par);
  double deltat,energy;
  /*@\morecode@*/
} ;  

MD_environment::MD_environment(int argc,char *argv[],MD_parameters& par) :
  Environment(arg,arv,par)
{
  deltat=par->value("deltat").as<double>();
  persist.reg_var("MD_environment.deltat",&deltat);
  persist.reg_var("MD_environmnet.energy",&energy);
  /*@\morecode@*/
}
\end{lstlisting}

We now inherit from \lstinline!Simulation! to define the simulation
step and the appropriate inizialization:
\begin{lstlisting}
class MD_simulation : public Simulation {
public:
  MD_simulation(MD_environment&,MD_configuration&);
  void step();
private:
  MD_environtment env;
  MD_configuration conf;
} ;

MD_simulation::MD_simulation(MD_environment& e,MD_configuration& c) :
  Simulation(e,c), env(e), conf(c)
{
  /*@\pseudo{compute initial energy}@*/
  /*@\pseudo{substract center-of-mass motion}@*/
  /*@\morecode@*/
}

void MD_simulation::step()
{
  /*@\pseudo{compute forces for configuration \emph{conf}}@*/
  /*@\pseudo{perform Verlet step of \emph{conf}}@*/
}
\end{lstlisting}

Additional logging (e.g.\ periodically writing the energy) and
observation (e.g. periodically saving the configuration to obtain a
trajectory) can be added to \lstinline!Simulation! and deriving from
\lstinline!Observable!.  Finally, one writes the \lstinline!create_xx!
functions that create the configuration, environment, and simulation
objects and return pointers to them (see listing~\ref{glsim-main}),
e.g.:
\begin{lstlisting}
Environment *create_environment(int argc,char *argv[],Parameters *par)
{
  return new MD_environment(argc,argv,*dynamic_cast<MD_parameters*>(par));
}
\end{lstlisting}

This code is then linked with \glsim\ and the \glsim-provided main to
produce a working \ac{MD} simulation. Based on the author's
experience, it is estimated that this new code amounts to between 300
and 600 source lines, to be compared with 2200+ lines in \glsim\
(providing command line parsing, configuration file parsing,
self-describing environment files and orderly interruption through
Unix signals). About half of the new code will be concerned with i/o
of configurations.

\subsection{Writing \lstinline!step()! in another language}

Although the definition of \lstinline!class MD_simulation! must
include the function \lstinline!step()! (otherwise the class would
remain abstract), it could be just a wrapper that calls routines in
another language, if that is convenient. The details of building a
mixed-language program depend on the language and platform (operating
system, linker, compiler) and can be rather tedious
\cite{algorithm:burow95}. However, C++ can be easily linked together
with C and FORTRAN (in most platforms). Linking with C is directly
supported by the standard through the 
\lstinline!extern "C" {!$\ldots$\lstinline!}!
construct. FORTRAN can be linked easily with
the aid of the \lstinline!cfortran.h! header \cite{cfortran}, which
supports a large number of compilers and linkers. To continue the
above example, if one wishes to use FORTRAN routines to compute the
forces and to perform the Verlet step, the above \lstinline!step()!
would look something like this:
\begin{lstlisting}
#include "cfortran.h"

PROTOCCALLSFFUN3(FORCE,force,DOUBLEVV,DOUBLEVV,INTV)
PROTOCCALLSFFUN3(VERLET,verlet,DOUBLEVV,DOUBLEVV,DOUBLEVV)

#define force(r,a,t) CCALLSFFUN3(FORCE,force,DOUBLEVV,DOUBLEVV,INTV,r,a,t)
#define verlet(r,v,a) CCALLSFFUN3(VERLET,verlet,DOUBLEVV,DOUBLEVV,DOUBLEVV,r,v,a)

void MD_simulation::step()
{
  force(conf.r,conf.a,conf.type);
  verlet(conf.r,conf.v,conf.a);
}
\end{lstlisting}

\lstinline!cfortran.h! also supports calling C or C++ from
FORTRAN. For details we refer to the \lstinline!cfortran.h!
documentation \cite{cfortran}.

\section{Final remarks and how to obtain \glsim}

\label{sec:conc}
We have described a library designed around an abstract definition of
a simulation, understood in a very general way, and built using
information hiding to provide convenient modules isolating
implementation details from the user. It is expected that it will be
useful to the developer of a program whose task can be described with
the basic simulation algorithm, helping him/her to fulfill most or all
of the requirements of a good simulation program with minimum effort.

\glsim\ is being used in actual research projects, and is under
development. Features (in particular checkpointing support) will be
added in the future. Also, since the design is open to addition of
more specialized modules, it is expected that the number of classes
will grow with modules adding support for more specific simulations
and for data analysis. It is hoped that the design of \glsim\ will
encourage its users to write modular, reusable code that can
eventually contribute to the growth of \glsim. The author has written
code for manipulating off-lattice configurations and trajectories
(along the lines sketched in sec.~\ref{sec:example}). This code is
useful for a wide variety of situations, including analysis programs
outside the simulation itself, and it is planned to add this to
\glsim\ as soon as the interface is polished. Hopefully others will
start using \glsim\ to later become contributors as well.

The \glsim\ source code is distributed under the GNU General Public
License version 3. It can be downloaded at no cost from SourceForge
(http://sourceforge.net/projects/glsim).

\section*{Acknowledgements}

I thank G.~Baglietto, M.~Carlevaro, E.~Loscar, and P.~Verrocchio for
critical reading of the manuscript. I also gratefully acknowledge many
discussions on numerical simulation issues with the aforementioned
colleagues and with E.~Albano, B.~Coluzzi, J.~R.~Grigera,
V.~Mart\'\i{}n-Mayor, and
G.~Parisi.

\bibliographystyle{model1-num-names}
\bibliography{glsim}

\end{document}